\documentclass[usenatbib]{mn2e}
\usepackage{graphicx,tabularx,amsmath,amssymb,upgreek,float}
\numberwithin{equation}{section}
\title[The shapes of cores in Ophiuchus]{The intrinsic shapes of starless cores in Ophiuchus}
\author[O. Lomax, A. P. Whitworth and A. Cartwright]{O. Lomax\thanks{E-mail: oliver.lomax@astro.cf.ac.uk}, A. P. Whitworth and A. Cartwright\\
School of Physics and Astronomy, Cardiff University, Cardiff CF24 3AA}

\begin{document}
\pagerange{\pageref{firstpage}--\pageref{lastpage}} \pubyear{2013}
\maketitle
\label{firstpage}

\begin{abstract}
Using observations of cores to infer their intrinsic properties requires the solution of several poorly constrained inverse problems. Here we address one of these problems, namely to deduce from the projected aspect ratios of the cores in Ophiuchus their intrinsic three-dimensional shapes. Four models are proposed, all based on the standard assumption that cores are randomly orientated ellipsoids, and on the further assumption that a core's shape is not correlated with its absolute size. The first and simplest model, {\bf M1}, has a single free parameter, and assumes that the relative axes of a core are drawn randomly from a log-normal distribution with zero mean and standard deviation $\sigma_{_{\rm O}}$. The second model, {\bf M2a}, has two free parameters, and assumes that the log-normal distribution (with standard deviation $\sigma_{_{\rm O}}$) has a finite mean, $\mu_{_{\rm O}}$, defined so that $\mu_{_{\rm O}}<0$ means elongated (prolate) cores are favoured, whereas $\mu_{_{\rm O}}>0$ means flattened (oblate) cores are favoured. Details of the third model ({\bf M2b}, two free parameters) and the fourth model ({\bf M4}, four free parameters) are given in the text. Markov chain Monte Carlo sampling and Bayesian analysis are used to map out the posterior probability density functions of the model parameters, and the relative merits of the models are compared using Bayes factors. We show that {\bf M1} provides an acceptable fit to the Ophiuchus data with $\sigma_{_{\rm O}}\approx0.57\pm0.06$; and that, although the other models sometimes provide an improved fit, there is no strong justification for the introduction of their additional parameters.
\end{abstract}


\vspace{0.5cm} 

\section{Introduction}%

{Cores are dense concentrations of interstellar matter in star-forming molecular clouds. Cores that are gravitationally bound are termed prestellar, and are expected to collapse to form stars. An individual prestellar core is normally assumed to spawn a single system (i.e. a single star, binary or multiple), or at most a small number of stars} \citep*[e.g.][]{AWB93,AWB00,HW13}. {There have been many attempts to simulate numerically the formation of stars in cores} \citep*[e.g.][]{B98,B00,HBB01,GW04,DCB04a,DCB04b,GWW04,GWW06,WBWNG09,WWG12}.

In subsequent papers, we will use SPH simulations to study (i) how an ensemble of cores produces a population of protostars and (ii) how the properties of these protostars compare with observed young stars. We will base the initial conditions for our simulations on observations of Ophiuchus, which is a convenient region for the study of star formation for two reasons. First, the Ophiuchus Main Cloud is relatively nearby, at a distance of $\sim\!130\,\mathrm{pc}$. This makes it possible to observe low-mass cores, and, in many cases, to resolve their spatial extent. Second, there is kinematic evidence which suggests that the cores in Ophiuchus are unlikely to interact with each other before they are through with forming stars \citep*{ABMP07}. This gives us some justification for modelling these cores as distinct isolated objects, albeit with the caveat that they are embedded in a substructured molecular cloud.

In this paper, we consider how to constrain the intrinsic three-dimensional shapes of the cores in Ophiuchus. There have been several previous models developed to fit the observed aspect ratios of cores, using both randomly oriented spheroids \citep[e.g.][]{MFGB91,R96} and randomly oriented ellipsoids \citep*[e.g.][]{JBD01,GW-TW02}. These models all invoke at least two free parameters. Here we introduce a model in which the intrinsic shapes of cores are characterised by just one free parameter. Using Markov Chain Monte Carlo sampling (MCMC), we generate a probability density function (PDF) for this parameter, based on observations of the cores in Ophiuchus by \citet*[][hereafter MAN98]{MAN98}, \citet*[][hereafter SSGK06]{SSGK06} and \citet*[][hereafter SNW08]{SNW-T08}. We also define three more complex models by introducing additional parameters. These sometimes provide a better fit, but we show that the improvement in fit does not justify the extra parameters.

In Section \ref{modelling} we introduce each of the models, and explain how we derive projected shapes. In Section \ref{observations} we review the observational data from MAN98, SSGK06 and SNW08. In Section \ref{bayesanalysis} we describe how we use Bayesian analysis to identify the best-fit model parameters, and the best models. In Section \ref{results} we present and discuss the results, and in Section \ref{conclusions} we summarise our conclusions.

\section{Modelling the shapes of cores}\label{modelling}%

We follow the convention of approximating core shapes with ellipsoids having semi-axes $A$, $B$ and $C$, where $A\geq B\geq C$. Furthermore, we assume that the intrinsic shape of a core is uncorrelated with its absolute size.

\subsection{Model {\bf M1}, one free parameter $(\sigma_{_{\rm O}})$}\label{SEC:M1}%

For the first model $\mathbf{M1}$, we generate a family of core shapes with only one free parameter: $\sigma_\textsc{o}$. Each individual shape is an ellipsoid with semi-axes
\begin{equation}
  \begin{split}
    A&=1\,,\\
    B&=\exp(\sigma_\textsc{o}\mathcal{G}_\textsc{b})\,,\\
    C&=\exp(\sigma_\textsc{o}\mathcal{G}_\textsc{c})\,.
  \end{split}
\label{EQN:M1}
\end{equation}
Here -- and in all further models -- $\mathcal{G}_\textsc{b}$ and $\mathcal{G}_\textsc{c}$ are random numbers drawn from a Gaussian distribution with zero mean and unit standard deviation. Once the semi-axes of a core have been generated, they are re-ordered so that $A\geq B\geq C$ and normalised so that $A=1$.

Increasing $\sigma_\textsc{o}$ increases the likelihood that the axes of a core have very disparate sizes, and hence the likelihood that the projected shape of the core has a small aspect ratio, $q$. Note that while this model \emph{can} produce oblate cores (i.e. $B\sim A$ and $C\ll B$) and prolate cores (i.e. $B\ll A$ and $C\sim B$), it \emph{does not} include a preference towards either shape. In general, the  individual shapes are simply triaxial (i.e. $A\neq B\neq C$).

Triaxial cores are likely to occur in the presence of turbulence. Even statistically isotropic turbulence will shock parcels of gas randomly along different directions, producing cores with random intrinsic aspect ratios. Furthermore, \citet*{LMS65} show that an ellipsoidal core that subsequently undergoes gravitational collapses tends to shrink fastest along its shortest axis; this will enhance any departures from spherical symmetry.

\subsection{Model {\bf M2a}, two free parameters $(\mu_{_{\rm O}},\sigma_{_{\rm O}})$}\label{SEC:M2a}%

For the second model $\mathbf{M2a}$, we generate a family of core shapes using two free parameters: $\mu_\textsc{o}$, $\sigma_\textsc{o}$. Each ellipsoid has semi-axes
\begin{equation}
  \begin{split}
    A&=1\,,\\
    B&=\exp(\mu_\textsc{o}+\sigma_\textsc{o}\mathcal{G}_\textsc{b})\,,\\
    C&=\exp(\mu_\textsc{o}+\sigma_\textsc{o}\mathcal{G}_\textsc{c})\,.
  \end{split}
\label{EQN:M2a}
\end{equation}
Unlike model $\mathbf{M1}$, model $\mathbf{M2a}$ explicitly includes the possibility of a preference for oblate or prolate cores. When $\mu_\textsc{o}\!\gtrsim\!\sigma_\textsc{o}$, cores tend to be oblate and when $\mu_\textsc{o}\!\lesssim\!-\,\sigma_\textsc{o}$, cores tend to be prolate.

Roughly axisymmetric shapes such as those provided by this model imply that the cores are, or have been, in some form of equilibrium. Oblate core shapes may occur if the self-gravity of a core is restricted by rotation \citep*[e.g.][]{KMH87} or by a poloidal magnetic field \citep*[e.g.][]{M76}. It is also possible that prolate cores may be the result of toroidal or helical magnetic fields \citep*[e.g.][]{T91,FP00a}.

\subsection{Model {\bf M2b}, two free parameters $(\sigma_{_{\rm B}},\sigma_{_{\rm C}})$}\label{SEC:M2b}%

For the third model $\mathbf{M2b}$, we generate a family of core shapes with two free parameters: $\sigma_\textsc{b}$ and $\sigma_\textsc{c}$. Each ellipsoid has semi-axes
\begin{equation}
  \begin{split}
    A&=1\,,\\
    B&=\exp(\sigma_\textsc{b}\mathcal{G}_\textsc{b})\,,\\
    C&=\exp(\sigma_\textsc{c}\mathcal{G}_\textsc{c})\,.
  \end{split}
\label{EQN:M2b}
\end{equation}
With model $\mathbf{M2b}$, if $\sigma_\textsc{b}\!\ll\! 1$ and $\sigma_\textsc{c}\!\gtrsim\! 1$, we produce an ensemble of oblate and prolate cores. While there is no strong physical justification for adopting this model, it provides a second relatively simple way of generating a two-parameter family of ellipsoidal shapes.

\subsection{Model {\bf M4}, four free parameters $(\mu_{_{\rm B}},\sigma_{_{\rm B}},\mu_{_{\rm C}},\sigma_{_{\rm C}})$}\label{SEC:M4}%

For the fourth and final model $\mathbf{M4}$, we generate a family of core shapes with four free parameters: $\mu_\textsc{b}$, $\mu_\textsc{c}$, $\sigma_\textsc{b}$ and $\sigma_\textsc{c}$ . Each ellipsoid has semi-axes
\begin{equation}
  \begin{split}
    A&=1\,,\\
    B&=\exp(\mu_\textsc{b}+\sigma_\textsc{b}\mathcal{G}_\textsc{b})\,,\\
    C&=\exp(\mu_\textsc{c}+\sigma_\textsc{c}\mathcal{G}_\textsc{c})\,.
  \end{split}
\label{EQN:M4}
\end{equation}
With model $\mathbf{M4}$, if $|\mu_\textsc{b}|\!>\!\sigma_\textsc{b}$ and $|\mu_\textsc{c}|\!>\!\sigma_\textsc{c}$ we produce cores with roughly the same triaxial shape. This shape will be much more varied between cores if $|\mu_\textsc{b}|\!\lesssim\!\sigma_\textsc{b}$ or $|\mu_\textsc{c}|\!\lesssim\!\sigma_\textsc{c}$.

As with model {\bf M1}, model $\mathbf{M4}$ implies that the core shapes have been perturbed by turbulence. The parameters of these models can also be compared to results by \citet{GW-TW02} and \citet{JBD01}. However, we note that it is difficult to imagine why turbulence would produce a population of cores all with roughly the same shape.

\subsection{Projecting an arbitrarily oriented ellipsoid}%

We define a Cartesian co-ordinate system in which the $x$-axis is aligned along $A$, the $y$-axis along $B$, and the $z$-axis along $C$. To observe this core from an arbitrary direction, given by polar angles $(\theta,\phi)$, we set
\begin{eqnarray}
\theta&=&\cos^{-1}(2{\cal R}_\theta-1)\,,\\
\phi&=&2\pi{\cal R}_\phi\,,
\end{eqnarray}
where ${\cal R}_\theta$ and ${\cal R}_\phi$ are random numbers drawn from a uniform distribution on the interval $(0,1)$. The aspect ratio of the core is then given by
\begin{eqnarray}
q&=&\sqrt{\frac{\alpha+\gamma-\sqrt{(\alpha-\gamma)^2+\beta^2}}{\alpha+\gamma+\sqrt{(\alpha-\gamma)^2+\beta^2}}}
\end{eqnarray}
where 
\begin{eqnarray}
\alpha\!&\!=\!&\!(A^2\cos^2(\phi)+B^2\sin^2(\phi))\cos^2(\theta)+C^2\sin^2(\theta),\\
\beta\!&\!=\!&\!(B^2-A^2)\cos(\theta)\sin(2\phi),\\
\gamma\!&\!=\!&\!A^2\sin^2(\phi)+B^2\cos^2(\phi)
\end{eqnarray}
\citep[see][]{B85}.

\begin{figure}[H]
   \centering
   \includegraphics[width=\columnwidth]{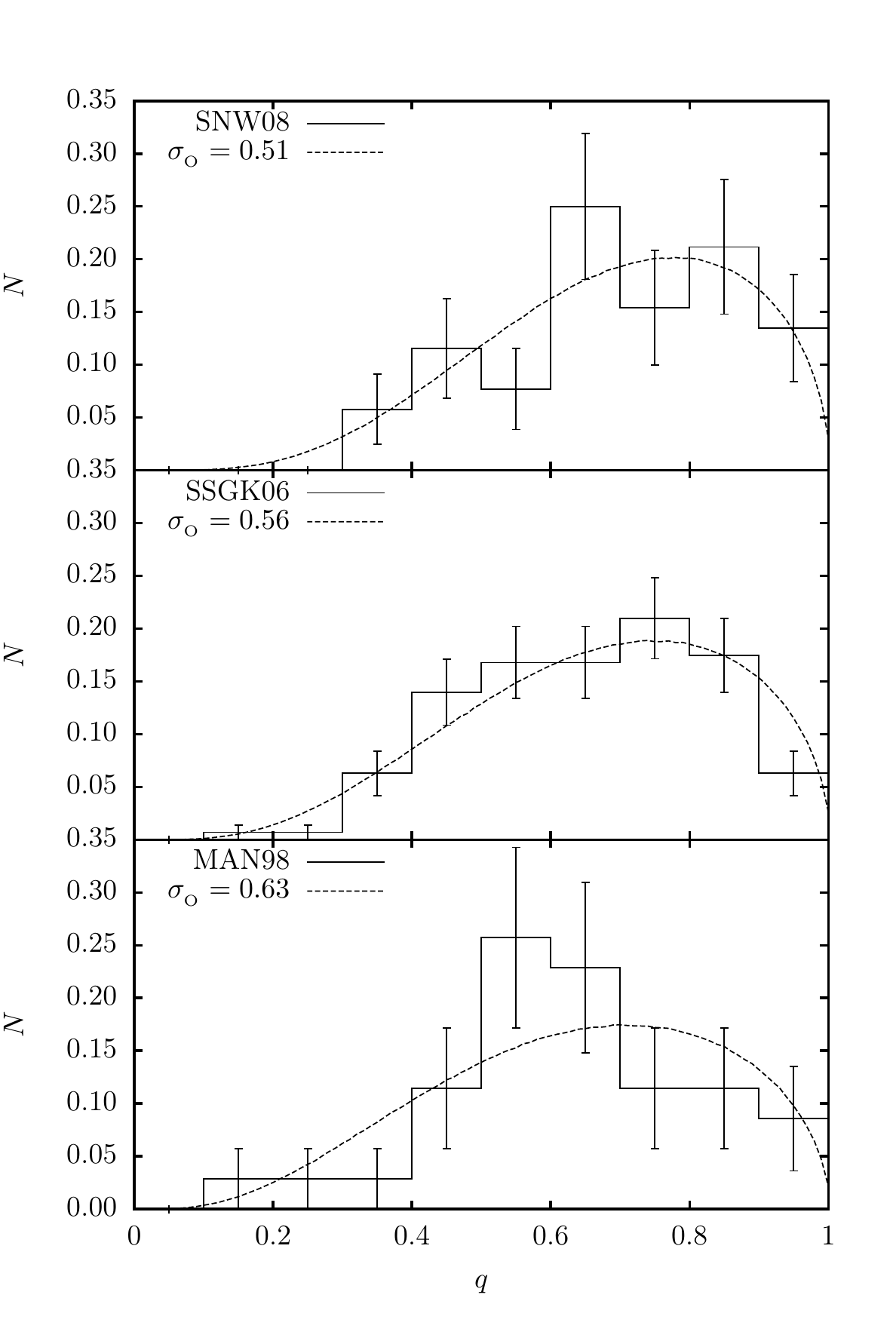}
   \caption[Aspect ratio fit from {\bf M1}.]{The histograms represent the distributions of aspect ratio obtained by SSGK06 (top), MAN98 (middle) and SNW08 (bottom), with $\sqrt{N}$ errors. The dashed lines represent the best fits obtained with {\bf M1}.}
   \label{aspect_data1}
\end{figure}

\section{Observational data}\label{observations}

{We apply the shape fitting analysis to observations of starless cores in Ophiuchus by MAN98, SSGK06 and SNW08. SNW08 and MAN98 present measurements of cores within the Ophiuchus Main Cloud. They conclude that most of these cores are probably prestellar, since their masses estimated from submillimetre continuum emission are comparable to or greater than their Jeans masses estimated from size and radial velocity dispersion. SSGK06 observe cores over a somewhat larger area, in and around the Ophiuchus Main Cloud. They note that many of the cores surrounding the main cloud have low surface densities and therefore may not be prestellar. In this section we give a brief overview of the observations.

\subsection{MAN98}%

MAN98 map dust emission at $1.3\,\mathrm{mm}$ using the MPIfR bolometer array on the IRAM $30\,\mathrm{m}$ telescope. The beam size is $11''$, corresponding to $1400\,{\rm AU}$ at Ophiuchus, and they are sensitive to hydrogen column-densities $N\!>\!10^{22}\,{\rm H}\,{\rm cm}^{-2}$. Their map covers an area of $\sim\!480\,\mathrm{arcmin}^2$ and includes the Oph-A, Oph-B1, Oph-B2, Oph-C, Oph-D, Oph-E and Oph-F clumps.\footnote{Here we adopt the nowadays near-universal convention that small condensations -- that might be prestellar -- are `cores', and more extended, diffuse, amorphous structures are `clumps'. MAN98 use the opposite notation, so the entities with which we are here concerned are referred to as `clumps' in their paper.} MAN98 use a multi-scale wavelet analysis to identify cores. They then subtract the background, and, where this is possible, fit the remaining intensity with a 2D ellipsoidal Gaussian, convolved with the beam. The dimensions of a core are then the FWHMs along the two principal axes of the fitted 2D ellipsoidal Gaussian. Where this fitting is not possible, the core is labelled unresolved, and these cores are not used in our analysis. The 36 starless cores used in our analysis have masses between $\sim\!0.1\,{\rm M}_{_\odot}$ and $\sim\!3\,{\rm M}_{_\odot}$, and dimensions between $\sim\!1000\,{\rm AU}$ and $\sim\!20000\,{\rm AU}$. MAN98 do not give the orientations of the resolved cores.

\subsection{SSGK06}%

SSGK06 map dust emission at $1.2\,\mathrm{mm}$ over a large area of sky (more than a square degree) around Ophiuchus, using the SIMBA bolometer array on the SEST telescope. The beam size is $24''$, corresponding to $3100\,{\rm AU}$ at Ophiuchus, and they are sensitive to hydrogen column-densities $N\!>\!4\times10^{21}\,{\rm H}\,{\rm cm}^{-2}$. They extract cores using wavelet decomposition and \emph{Clumpfind} \citep*{WGB94}. For 111 starless cores having peak intensities at least three times greater than the background noise, they compute the FWHM along the major and minor axes, and the orientation of the major axis, using the moments of the intensity map. The listed values are not deconvolved with the beam, but in general they are several times larger than the beam, and therefore the resulting reduction in projected aspect ratios should only affect significantly the smallest cores. We also do not correct for this, and the fact that there is no noticeable excess of very circular cores ($q\sim 1$) in the SSGK06 distribution indicates that this is sensible. We have therefore used all 111 starless cores from SSGK06 in our analysis. They have masses between $\sim\!0.02\,{\rm M}_{_\odot}$ and $\sim\!6\,{\rm M}_{_\odot}$, and dimensions between $\sim\!2400\,{\rm AU}$ and $\sim\!40000\,{\rm AU}$.

\begin{figure}
   \centering
   \includegraphics[width=0.6\columnwidth,angle=90]{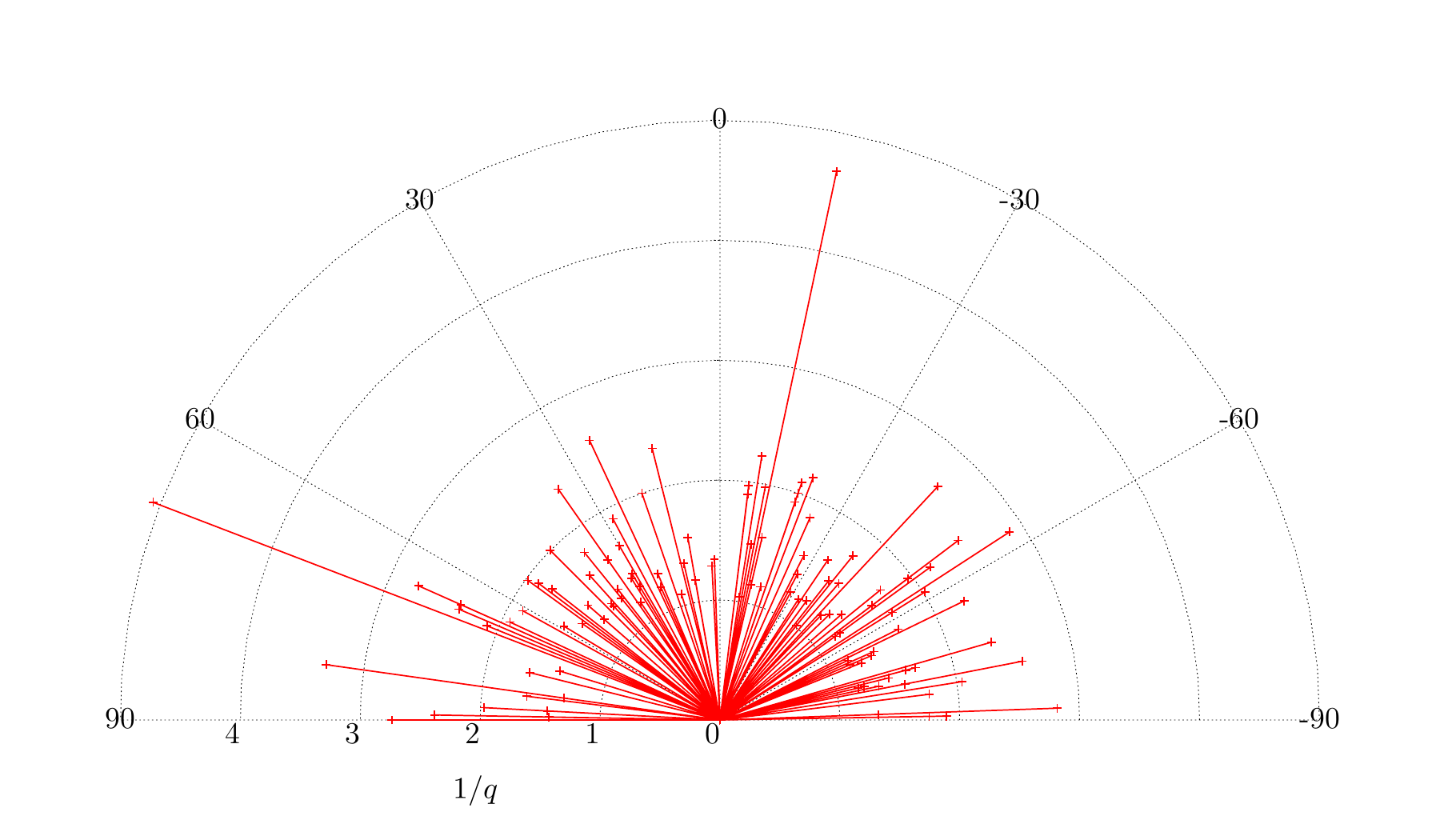}
   \includegraphics[width=\columnwidth]{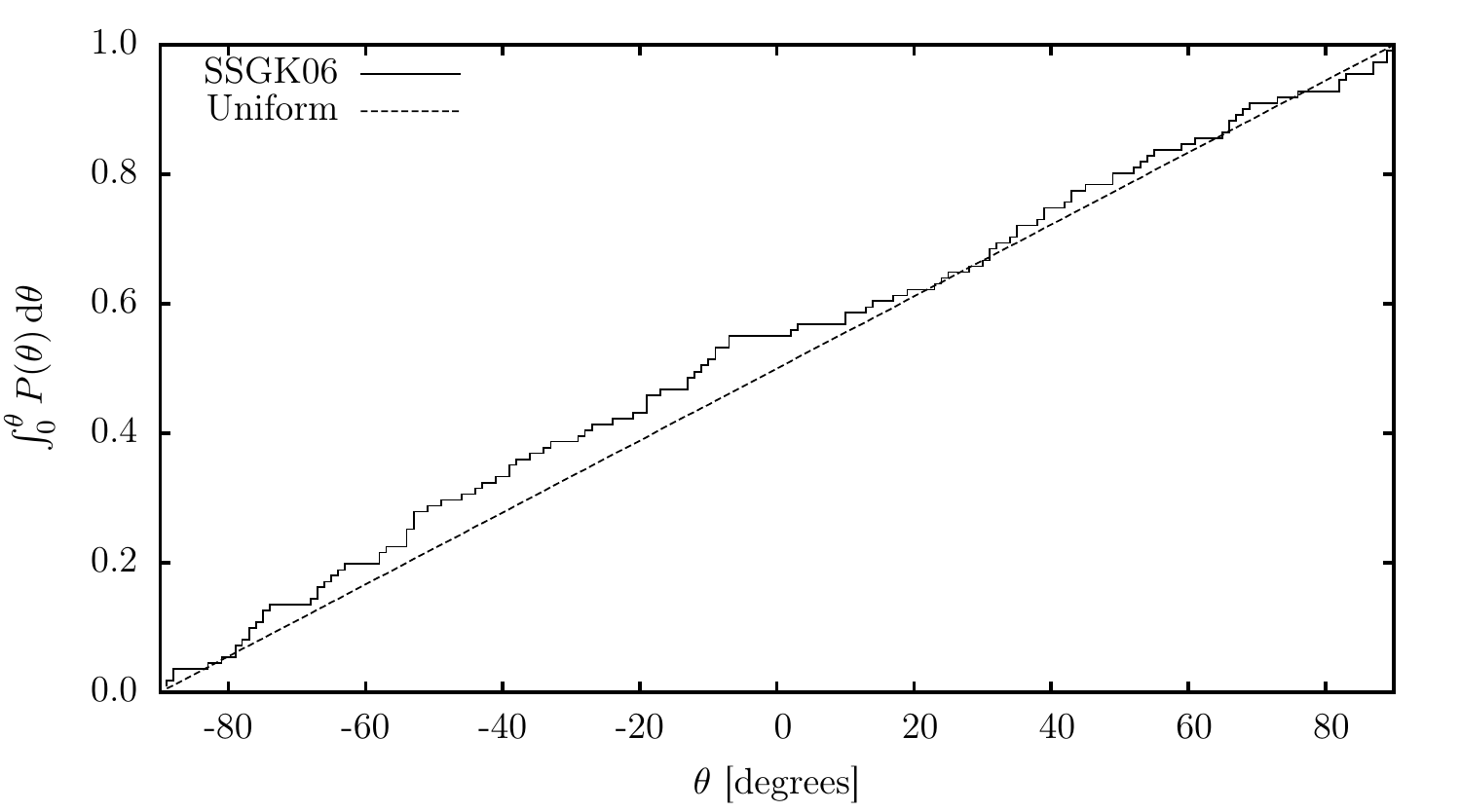}
   \caption[The distribution of orientations of SSGK06 cores.]{(a) A polar diagram showing the directions along which the SSGK06 cores are elongated; the length of each vector is inversely proportional to the aspect ratio, $q^{-1}$, so that more elongated cores have longer vectors. (b) The cumulative distribution of directions.}
   \label{FIG:ORIENTATIONS}
\end{figure}

\begin{figure}
   \centering
   \includegraphics[width=\columnwidth]{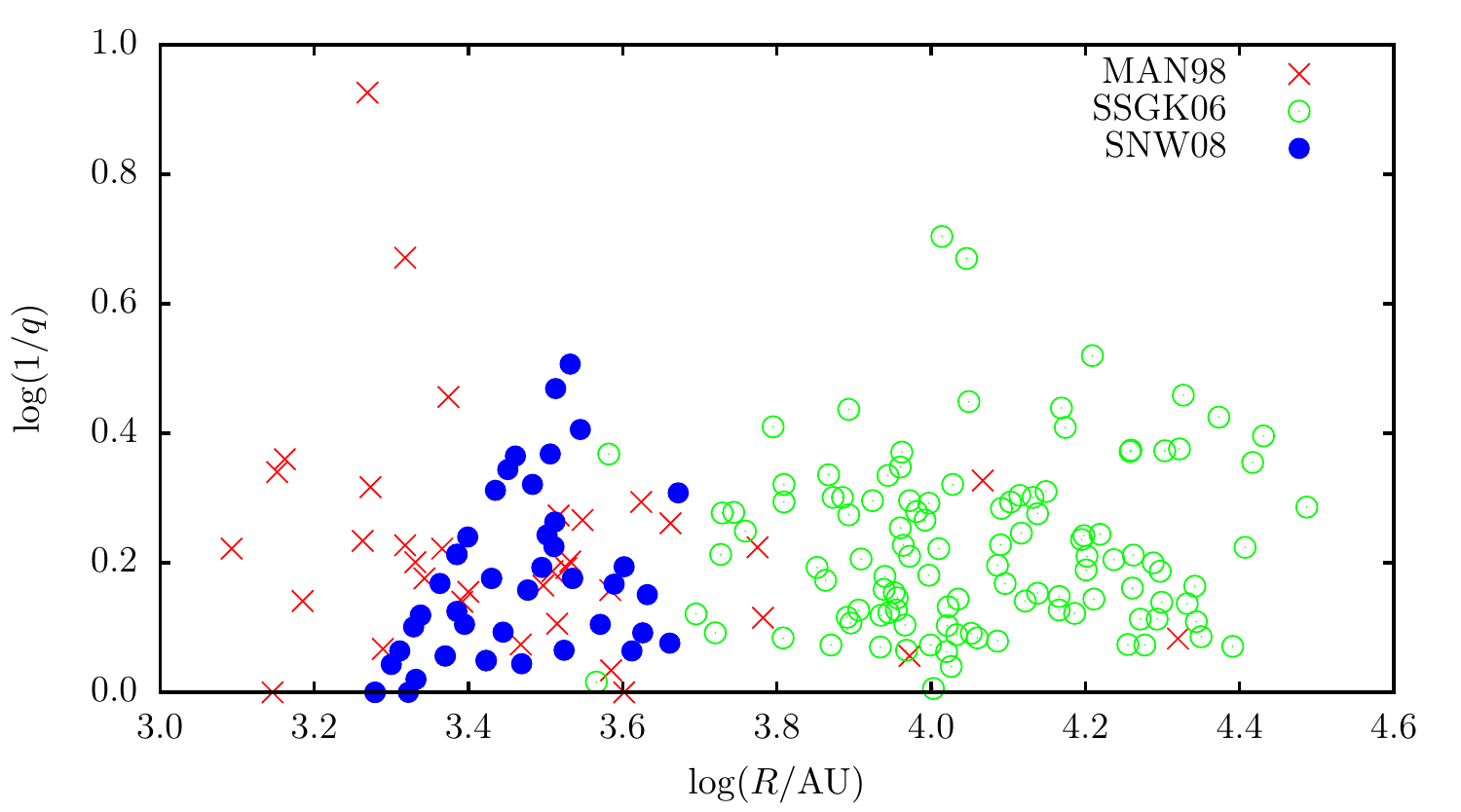}
   \caption[Shape against size.]{The inverse aspect ratios of cores, $q^{-1}$, plotted against their mean radii, $\bar{r}\!=\!(ab)^{1/2}$. Crosses represent the MAN98 data, open circles represent the SSGK06 data, and filled circles represent the SNW08 data.}
   \label{FIG:SHAPE/SIZE}
\end{figure}

\subsection{SNW08}%
\label{snw08data}

SNW08 analyse SCUBA observations of $850\,\mu{\rm m}$ dust emission from Ophiuchus, retrieved from the Canadian Astronomy Data Centre's JCMT data archive \citep*{TJEC97}; the beam size is $14''$, corresponding to $1900\,{\rm AU}$ at Ophiuchus, and they are sensitive to hydrogen column-densities $N\!>\!3\times10^{21}\,{\rm H}\,{\rm cm}^{-2}$. The data cover an area of $\sim\!700\,\mathrm{arcmin}^2$ that includes Oph-A, Oph-B1, Oph-B2, Oph-C, Oph-E Oph-F and Oph-J. Sources with peak brightness at least 5 times the background noise are delineated at 3 times the background noise. 52 cores are adjudged to be resolved, but are not deconvolved with the beam, and are fitted with ellipses, by eye, to obtain major and minor axes. We note that nearly 40\% of the cores have minor axes comparable with the beam size, which could be why this data-set has (i) a somewhat higher fraction of cores with large aspect ratios ($0.9\!\leq\!q\!<\!1$) and (ii) the smallest best-fit value of $\sigma_{_{\rm O}}$ (see Fig. \ref{m1params}). However, this is a small effect, and we do not correct for it. The 52 starless cores used in our analysis have masses between $\sim\!0.01\,{\rm M}_{_\odot}$ and $\sim\!6\,{\rm M}_{_\odot}$, and dimensions between $\sim\!1900\,{\rm AU}$ and $\sim\!6700\,{\rm AU}$. Again, SNW08 do not give the orientations of their cores.

\subsection{Caveats}%

Fig. \ref{aspect_data1} shows the distributions of aspect ratio for the three data-sets. However, because real cores are not isolated, for example they are often embedded in filaments, and because their internal iso-density surfaces do not conform to symmetrically nested, equal-eccentricity ellipsoids, the observational data must be viewed with caution. First, the observed intensity maps are likely to have contributions from other regions along the line of sight. Even after background subtraction, this may confuse the inferences made regarding the core under investigation. Second, the temperature in a core will not be uniform \citep[e.g.][]{SWW07}, and so monochromatic intensity is not necessarily an accurate proxy for dust column-density. Third, the dust may not be co-extensive with the gas; there is some observational evidence, as well as theoretical reasoning, suggesting that the gas is less centrally concentrated than the dust \citep[e.g.][]{PC06,WB02}. Fourth, a core that is embedded in a filament, and possibly gaining mass from the filament, does not have a well defined boundary. If the shape and orientation are measured using a low-intensity contour, they may be corrupted by the filament; this suggests that the 2D Gaussian fitting technique, which better reflects the central regions of the core, is probably superior to the other techniques.

\subsection{Systematic effects}%

In addition, there may be systematic effects that need to be taken into the reckoning. 

{\sc Environmental considerations.} Ophiuchus is located at the boundary of the Upper Scorpius OB Association, and the overall molecular cloud has a very filamentary shape, suggesting that it may have been compressed -- indeed, its star formation may have been triggered -- by the expansion of H{\sc ii} Regions, stellar winds and supernovae from this association. The possibility then exists that the cores in Ophiuchus are not randomly oriented, but have a preferred orientation, with their long axes either perpendicular or parallel to the line pointing towards Upper Sco. To explore this possibility, Fig. \ref{FIG:ORIENTATIONS}(a) shows a polar diagram of the orientation vectors from SSGK06, with the length of each vector being inversely proportional to the corresponding aspect ratio. Fig. \ref{FIG:ORIENTATIONS}(b) shows the cumulative distribution of the associated polar angles. It is clear from these figures that there is no strong preference for a particular orientation, and therefore there is no evidence here to challenge our assumption of random orientations.

{\sc Evolutionary considerations.} The assumption that the shape of a starless core is independent of its size is largely dictated by expediency: we have too few data-points to significantly constrain a bivariate distribution. If a core condenses out of a thin layer, shell or filament, one might expect it to become more spherical as it evolves. Conversely, if a core enjoys significant large-scale magnetic of rotational support, or if its condensation approaches freefall, one might expect it to become less spherical as it shrinks. Fig. \ref{FIG:SHAPE/SIZE} shows the logged inverse aspect ratios, $q^{-1}$, for the three data-sets, plotted against logged mean radius, $R\!=\!(ab)^{1/2}$. The Pearson  correlation coefficients are $\rho_{_{\rm MAN98}}=-0.24$, $\rho_{_{\rm SSGK06}}=0.08$, and $\rho_{_{\rm SNW08}}=0.40$. There does not appear to be any strong evidence for a correlation between shape and size within the MAN98 and SSGK06 data. There is some positive correlation between elongation and size in the SNW08 data due to the resolution effects, as mentioned in Section \ref{snw08data}\,.}

\section{Bayesian analysis}\label{bayesanalysis}%

\begin{figure}
   \centering
   \includegraphics[width=\columnwidth]{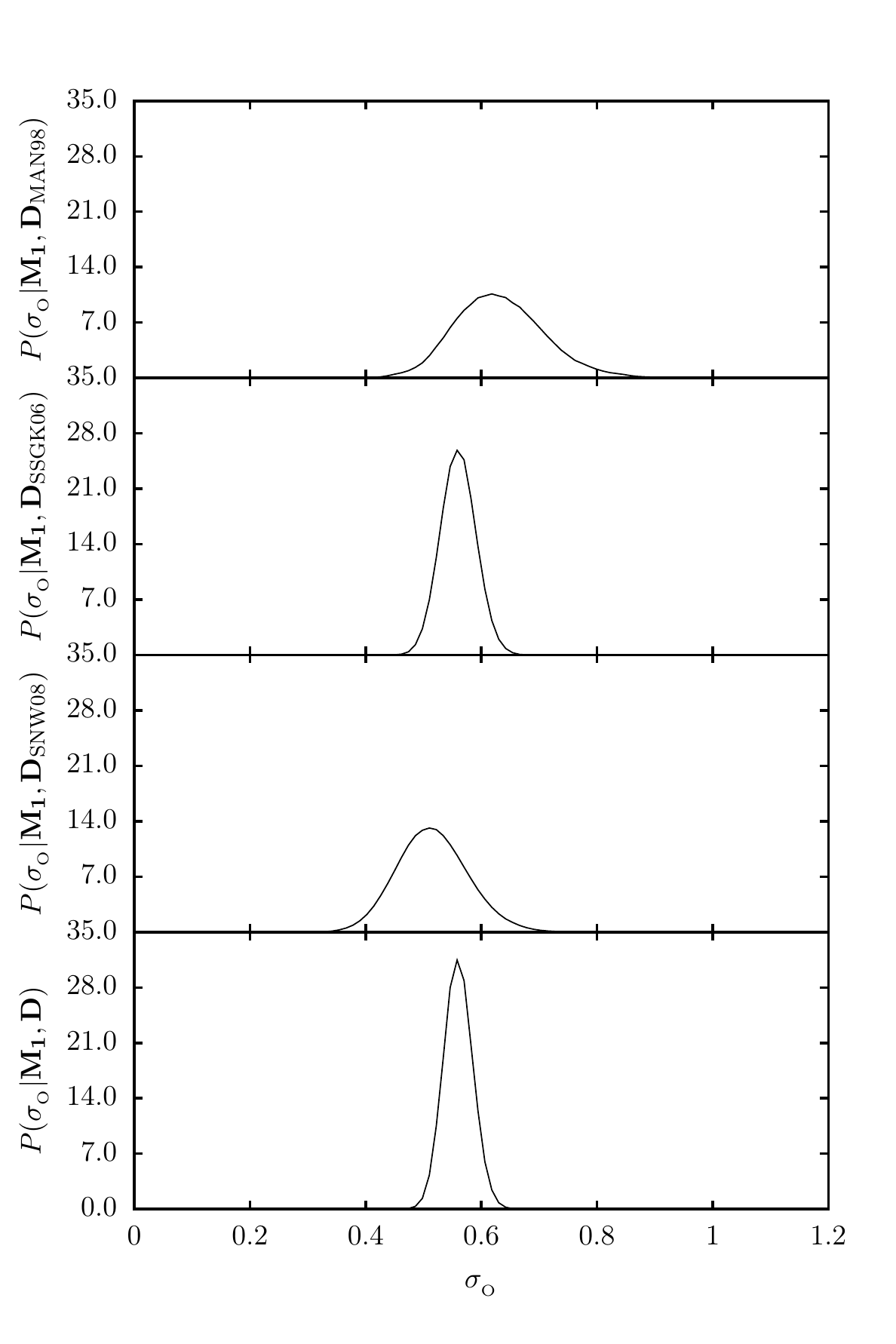}
   \caption[Posterior PDFs for $\sigma_{_{\rm O}}$ in {\bf M1}.]{Posterior PDFs for $\sigma_{_{\rm O}}$ in {\bf M1}, from the MAN98, SSGK06 and SNW08 data. The bottom panel shows the product of all three PDFs, i.e. our inference of $\sigma_{_{\rm O}}$ given all three data sets. The length of the $x$-axis represents the prior range given in Equation (\ref{priorpdf1}).}
   \label{m1params}
\end{figure}

\begin{figure}
   \centering
   \includegraphics[width=0.8\columnwidth]{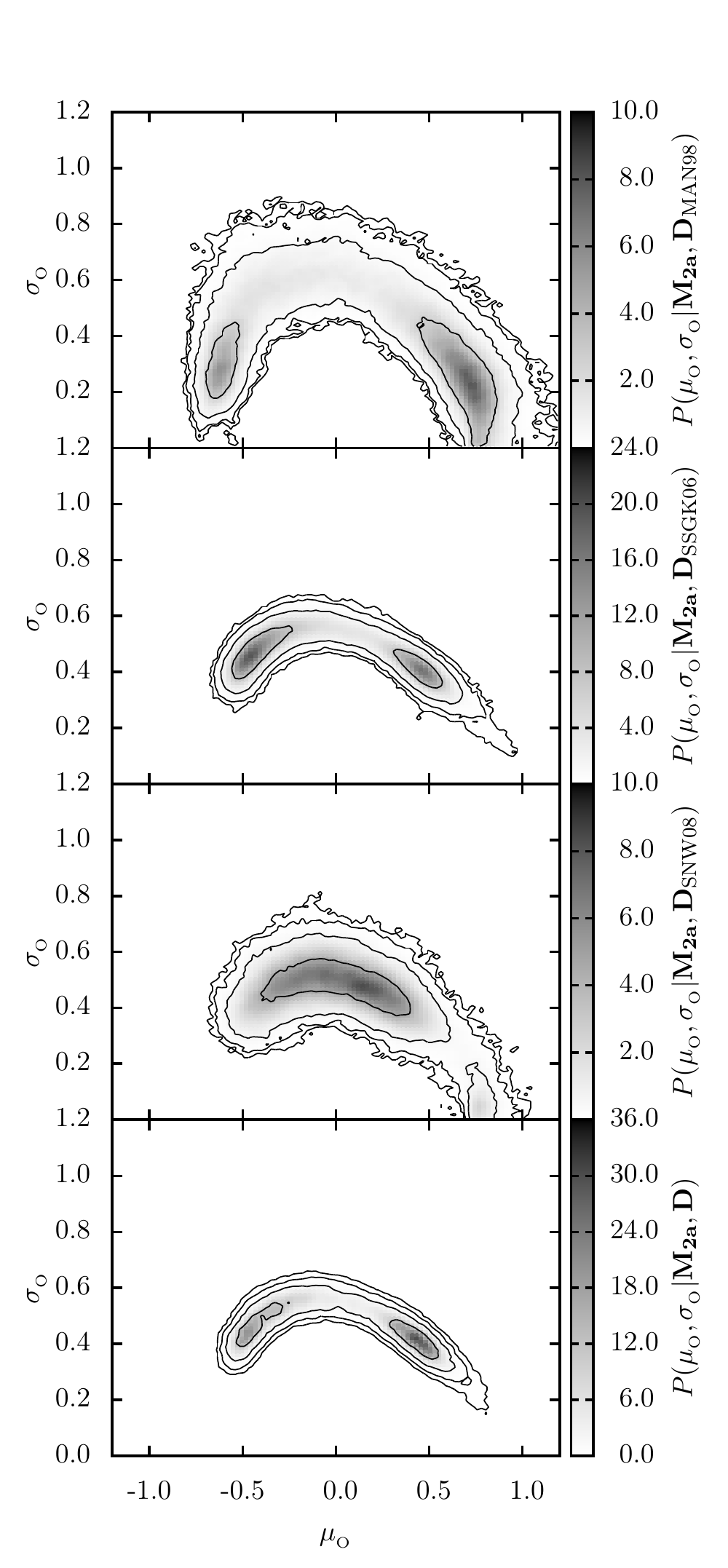}
   \caption[Posterior PDFs for $\mu_{_{\rm O}}$ and $\sigma_{_{\rm O}}$ in {\bf M2a}.]{Posterior PDFs for $\mu_{_{\rm O}}$ and $\sigma_{_{\rm O}}$ in {\bf M2a}, from the MAN98, SSGK06 and SNW08 data. The bottom panel shows the product of all three PDFs, i.e. our inference of $\mu_{_{\rm O}}$ and $\sigma_{_{\rm O}}$ given all three data sets. The contours outline the most likely 50\%, 95\%, 99.5\% and 99.95\% of the PDF. The size of the plotting area represents the prior ranges given in Equation (\ref{priorpdf2}).}
   \label{m2params}
\end{figure}

\begin{figure}
   \centering
   \includegraphics[width=0.8\columnwidth]{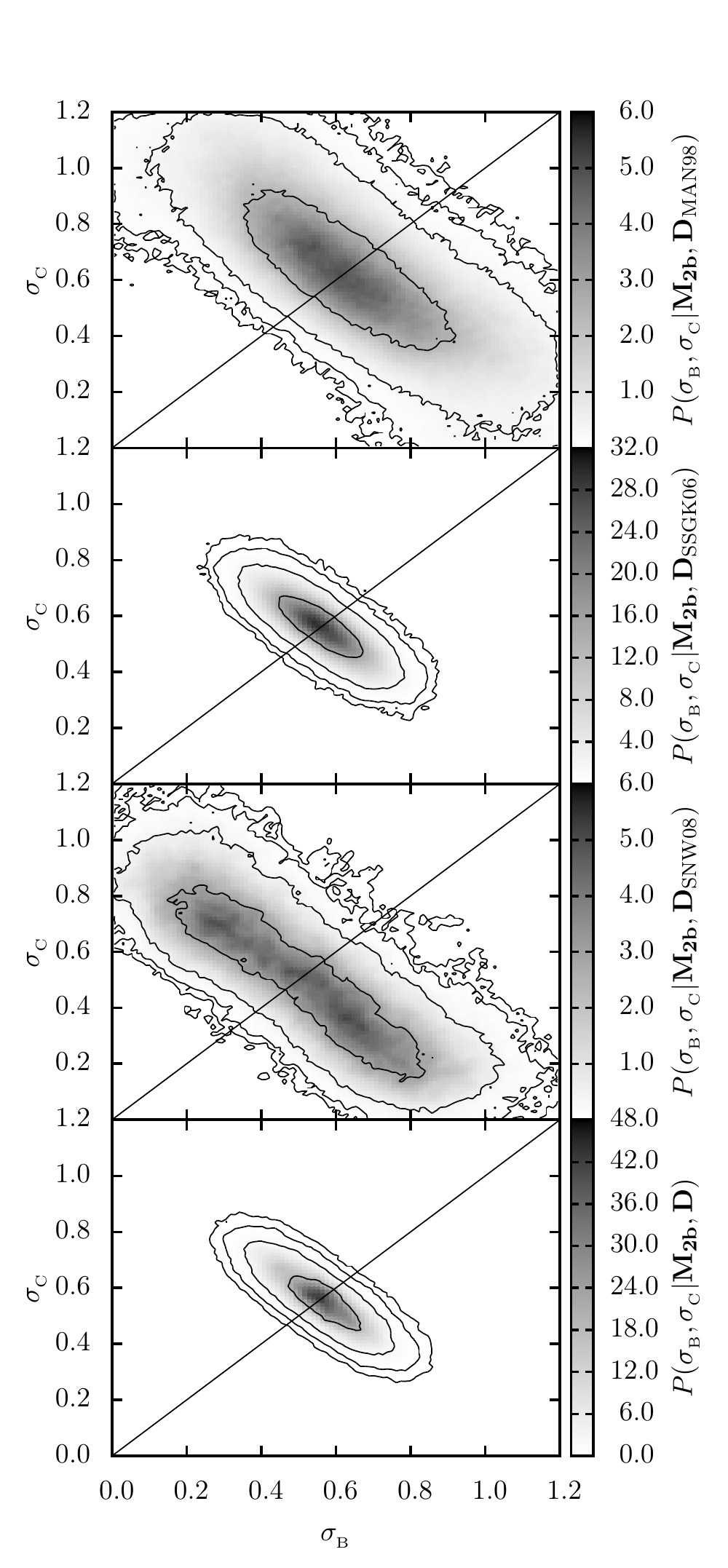}
   \caption[Posterior PDFs for $\sigma_{_{\rm B}}$ and $\sigma_{_{\rm C}}$ in {\bf M2b}.]{Posterior PDFs for $\sigma_{_{\rm B}}$ and $\sigma_{_{\rm C}}$ in {\bf M2b}, from the MAN98, SSGK06 and SNW08 data. The bottom panel shows the product of all three PDFs, i.e. our inference of $\sigma_{_{\rm B}}$ and $\sigma_{_{\rm C}}$ given all three data sets. The contours outline the most likely 50\%, 95\%, 99.5\% and 99.95\% of the PDF. The diagonal line represents $\sigma_{_{\rm B}}=\sigma_{_{\rm C}}$, about which the distribution should be symmetric. The size of the plotting area represents the prior ranges given in Equation (\ref{priorpdf3}).}
   \label{m3params}
\end{figure}

\begin{figure}
   \centering
   \includegraphics[width=0.8\columnwidth]{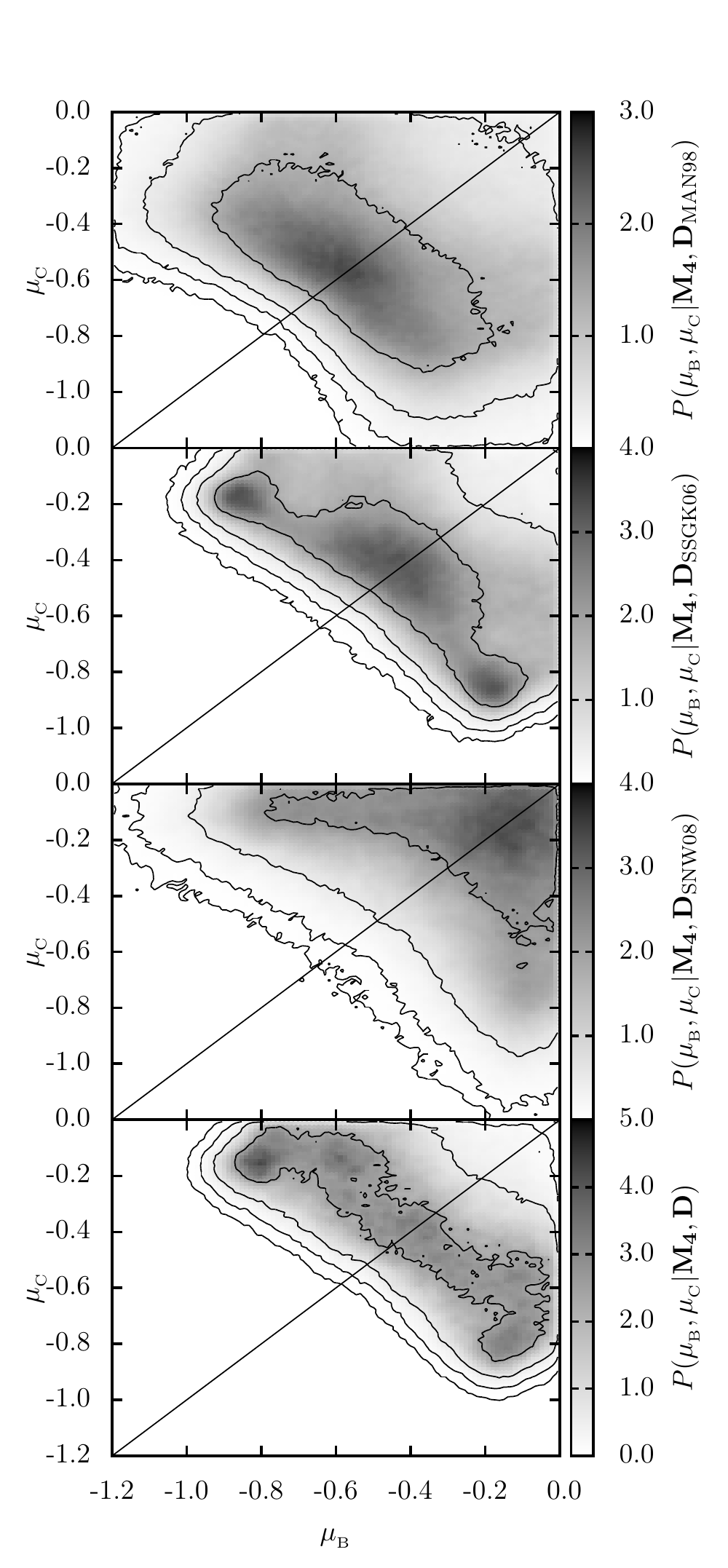}
   \caption[Posterior PDFs for $\mu_{_{\rm B}}$, $\mu_{_{\rm C}}$ in {\bf M4}.]{Posterior PDFs for $\mu_{_{\rm B}}$, $\mu_{_{\rm C}}$ in {\bf M4}, from the MAN98, SSGK06 and SNW08 data. $\sigma_{_{\rm B}}$ and $\sigma_{_{\rm C}}$ have been marginalized out, i.e. $P(\mu_{_{\bf B}},\mu_{_{\rm C}}|{\bf M4},{\bf D})=\iint P(\mu_{_{\rm B}},\mu_{_{\rm C}},\sigma_{_{\rm B}},\sigma_{_{\rm C}}|{\bf M4},{\bf D})\,{\rm d}\sigma_{_{\rm B}}\,{\rm d}\sigma_{_{\rm C}}$. The bottom panel shows the product of all three PDFs, i.e. our inference of $\mu_{_{\rm B}}$, $\mu_{_{\rm C}}$ given all three data sets. The contours outline the most likely 50\%, 95\%, 99.5\% and 99.95\% of the PDF. The diagonal line represents $\mu_{_{\rm B}}\!=\!\mu_{_{\rm C}}$, about which the distribution should be symmetric. The size of the plotting area represents the prior ranges given in Equation (\ref{priorpdf4}).}
   \label{m4params}
\end{figure}

\begin{figure}
   \centering
   \includegraphics[width=\columnwidth]{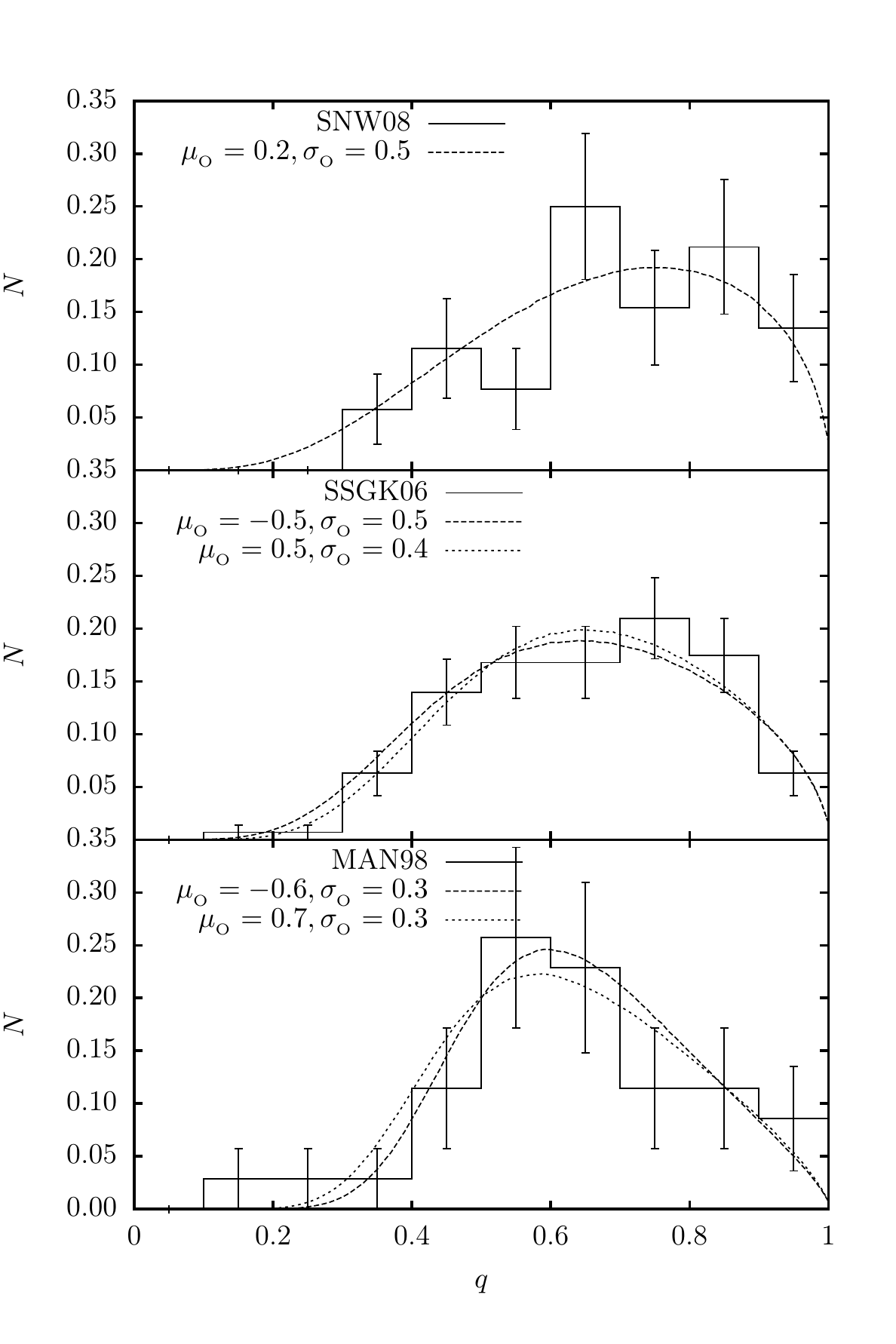}
   \caption[Aspect ratio fit from {\bf M2a}.]{As Figure \ref{aspect_data1}, but for {\bf M2a}.}
   \label{aspect_data2}
\end{figure}

We use Bayesian analysis to determine the best-fit parameters of the different models, and to quantify their relative strengths. When comparing model {\bf M} with parameters ${\bf x}\equiv\!(x_1,x_2,...)$ against observational data {\bf D}, Bayes' theorem states that
\begin{equation}
  P(\mathbf{x}|\mathbf{M},\mathbf{D})=\frac{P(\mathbf{D}|\mathbf{M},\mathbf{x})P(\mathbf{x}|\mathbf{M})}{P(\mathbf{D|M})}\,.
\label{bayes}
\end{equation}
Here $P(\bf{x}|\bf{M},\bf{D})$ is the posterior probability of {\bf x} given {\bf D}, $P(\bf{D}|\bf{M},\bf{x})$ is the likelihood of {\bf D} given {\bf x}, $P(\bf{x}|\bf{M})$ is the prior PDF of {\bf x} and $P(\bf{D|M})$ is the marginal likelihood over all values of {\bf x}, i.e.
\begin{equation}
  P(\bf{D|M})=\int_{\bf x}P(\bf{D}|\bf{M},\bf{x})P(\bf{x}|\bf{M})\,\rm{d}\bf{x}\,.
\label{marginal}
\end{equation}
As $P(\bf{D|M})$ is a constant, Equation (\ref{bayes}) simplifies to
\begin{equation}
  P(\bf{x}|\bf{M},\bf{D})\propto P(\bf{D}|\bf{M},\bf{x})P(\bf{x}|\bf{M})\,,
\end{equation}
where any generated posterior PDFs can be normalized to unity, post analysis.

\subsection{Prior PDF}\label{prior}%

When generating prior PDFs for the model parameters {\bf x} we assume that $P(\bf{x}|\bf{M})$ is finite and uniform within given limits, and zero outside them. This is to say, within credible limits, we impose no {\it a priori} preference for any specific {\bf x}.

For {\bf M1}, the single parameter $\sigma_{_{\rm O}}$ must be able to reproduce the maximum and minimum observed aspect ratios in the data, viz. $q_{_{\rm MAX}}\approx 1$ and $q_{_{\rm MIN}}\approx 0.3\;$ (over all three data sets, there are only two cores with $q\!<\!0.3$). Since the majority of aspect ratios delivered by {\bf M1} satisfy $q\!\gtrsim\!\exp(-\sigma_{_{\rm O}})$, we set $-\ln(q_{_{\rm MAX}})\!\leq\!\sigma_{_{\rm O}}\!\leq\!-\ln(q_{_{\rm MIN}}),\;$ i.e. $0\!\leq\!\sigma_{_{\rm O}}\!\leq\!1.2\,$.

For {\bf M2a} we set the range of $\mu_{_{\rm O}}$ to $-1.2\!\leq\!\mu_{_{\rm O}}\!\leq\!1.2$ so that a purely oblate or prolate population (i.e. one with $\sigma_{_{\rm O}}\!=\!0$) could reproduce the observed aspect ratios. We then assign $\sigma_{_{\rm O}}$ the same range as in {\bf M1}.

For both {\bf M2b} and {\bf M4} we assign $\sigma_{_{\rm B}}$ and $\sigma_{_{\rm C}}$ the same range as $\sigma_{_{\rm O}}$ in {\bf M1}. For {\bf M4} we assign $\mu_{_{\rm B}}$ and $\mu_{_{\rm C}}$ the same range as $\mu_{_{\rm O}}$ in {\bf M2a}.

For {\bf M4}, there are four degenerate regions of $(\mu_{_{\rm B}},\mu_{_{\rm C}})$ that can define all possible ellipsoidal shapes. To break this degeneracy, we adopt ranges $-1.2\!\leq\!\mu_{_{\rm B}}\!\leq\!0$ and $-1.2\!\leq\!\mu_{_{\rm C}}\!\leq\!0$ (strictly speaking, we should adopt $-1.2\!\leq\!\mu_{_{\rm C}}\!\leq\!\mu_{_{\rm C}}$, however this results in a triangular plotting area which is less convenient to plot than a square one). For $\sigma_{_{\rm B}}$ and $\sigma_{_{\rm C}}$, we assign the same range as in ${\bf M2b}$\,.

With these ranges, the normalised prior PDFs are:
\begin{equation}
  P(\sigma_{_{\rm O}}|\bf{M1})=\left\{\!
    \begin{array}{ll}
      \frac{1}{(1.2)} & \text{if } 0\!\leq\!\sigma_{_{\rm O}}\!\leq\!1.2,\\
      0 & \text{otherwise};
    \end{array} \right.
    \label{priorpdf1}
\end{equation}
\begin{equation}
  P(\mu_{_{\rm O}},\sigma_{_{\rm O}}|\bf{M2a})=\left\{\!
    \begin{array}{ll}
      \frac{1}{2(1.2)^2} & \text{if } -\!1.2\!\leq\!\mu_{_{\rm O}}\!\leq\!1.2\\
      &\text{and } 0\!\leq\!\sigma_{_{\rm O}}\!\leq\!1.2,\\
      0 & \text{otherwise};
    \end{array} \right.
    \label{priorpdf2}
\end{equation}
\begin{equation}
  P(\sigma_{_{\rm B}},\sigma_{_{\rm C}}|\bf{M2b})=\left\{\!
    \begin{array}{ll}
      \frac{1}{(1.2)^2} & \text{if }  0\!\leq\!\sigma_{_{\rm B}}\!\leq\!1.2\\
      &\text{and } 0\!\leq\!\sigma_{_{\rm C}}\!\leq\!1.2,\\
      0 & \text{otherwise};
    \end{array} \right.
    \label{priorpdf3}
\end{equation}
\begin{equation}
  P(\mu_{_{\rm B}},\mu_{_{\rm C}},\sigma_{_{\rm B}},\sigma_{_{\rm C}}|\bf{M_4})=\left\{\!
    \begin{array}{ll}
      \frac{1}{(1.2)^4} & \text{if }  -\!0\!\leq\!\mu_{_{\rm B}}\!\leq\!0\\
      &\text{and } -\!1.2\!\leq\!\mu_{_{\rm C}}\!\leq\!1.2\\
      &\text{and } 0\!\leq\!\sigma_{_{\rm B}}\!\leq\!1.2\\
      &\text{and } 0\!\leq\!\sigma_{_{\rm C}}\!\leq\!1.2,\\
      0 & \text{otherwise}\,.
    \end{array} \right.
\label{priorpdf4}
\end{equation}
Note that these prior PDFs \emph{will not} affect the inferences on specific parameter values as long as the posterior distribution of {\bf x} is within these credible limits. These priors \emph{will} affect the strength of specific models, as discussed further in Section \ref{selection}\,.

\subsection{Markov chain Monte Carlo sampling}%

For each observational data set, {\bf D}, we generate a histogram of aspect ratios. The histogram has ten bins ($k\!=\!1\,{\rm to}\,10$), evenly spaced between $q=0$ and $q=1$, and $O_k$ is the number of observed cores in bin $k$.

For a given model, {\bf M}, and a given choice of the associated free parameters, ${\bf x_i}$, we generate $10^4$ ellipsoids, and view each one from an arbitrary direction to determine its aspect ratio, $q$, as described in Section \ref{modelling}. The resulting $q$-values are then used to construct an equivalent histogram of expectation values, $E_j,\;(j\!=\!1\,{\rm to}\,10)$, normalised so that $\sum_j\{E_j\}\!=\!\sum_j\{O_j\}$. The likelihood of the observational data, {\bf D}, being reproduced by ${\bf M},{\bf x_i}$ is then
\begin{equation}
  P({\bf D}|{\bf M},{\bf x_i})=\exp\left\{-\,\sum\limits_{j=1}^{j=10}\frac{(O_j-E_j)^2}{2O_j}\right\}\,.
\label{chi2}
\end{equation}
We have assumed purely Poisson errors on the counts in each bin $O_j$ because error estimates for individual observed $q$-values are not available. Bins that have less than five counts are pooled together so that the Gaussian approximation to Poisson errors is valid.

To build a Markov Chain, we consider the observational values, $O_k$, from a particular data set, {\bf D}, and we invoke a particular model, {\bf M}. We pick a set of model parameters (${\bf x_0}$) in the middle of the ranges defined in Section \ref{prior}, and compute $P({\bf D}|{\bf M},{\bf x_0})$, as described in the preceding paragraph. We then build the chain by stepping from one set of model parameters to another, ${\bf x_0}\rightarrow{\bf x_1}\rightarrow{\bf x_2}\rightarrow{\bf x_3}...$. Each step, $\Delta{\bf x}\!=\!{\bf x_{i+1}-x_i}$ is drawn randomly from a Gaussian distribution centred on zero. The step is only made if
\begin{equation}
  \frac{P(\bf{D}|\bf{M},\bf{x_{i+1}})}{P(\bf{D}|\bf{M},\bf{x_i})}\geq\cal{R}_{_{\rm STEP}}\,,
\label{lratio}
\end{equation}
where ${\cal R}_{_{\rm STEP}}$ is a random number from a uniform distribution on the interval (0,1). Otherwise the step is rejected and a new step is drawn; this ensures that the points on the chain tend to concentrate in regions of high probability. The coefficients regulating the mean step size should be adjusted so that roughly half the steps are rejected. The first $10^3$ points on the chain are discarded, to remove any memory of the starting point. The subsequent $5\times 10^5$ points are used to identify the best-fit parameters and their uncertainties.

We have built a Markov Chain for each possible combination of the four models and the three data sets. The points on the chain are then used to determine the posterior PDFs of the model parameters. The results are presented in Figures \ref{m1params}, \ref{m2params}, \ref{m3params} and \ref{m4params}. The best fits obtained with {\bf M1} and {\bf M2a} are compared with the observations in Figures \ref{aspect_data1} and \ref{aspect_data2}.

\subsection{Model selection}\label{selection}%

Bayesian analysis can also be used to compare different models. Given a list of competing models, $\bf{M_1},\bf{M_2},\mathellipsis,\bf{M_n}$, the probability of a particular model, $\bf{M_k}$, is
\begin{equation}
   P(\bf{M_k}|\bf{D})=\frac{P(\bf{D}|\bf{M_k})P(\bf{M_k})}{P(\bf{D})}\,,
\end{equation}
where
\begin{equation}
   P(\bf{D})=\sum\limits_{k=1}^{k=n}P(\bf{D}|\bf{M_k})P(\bf{M_k})\,.
\end{equation}
To calculate $P(\bf{D}|\bf{M_k})$ we must marginalise each model's likelihood over its associated parameter space (see Eqn. \ref{marginal}). We evaluate this integral by organising the points on the associated Markov Chain into a balanced binary tree \citep{W09}. This has the effect of dividing the parameter space into cells, each of which contains a single point. Each point, $\bf{x_i}$, now has a likelihood (see Eqn. \ref{chi2}) and a volume of parameter space, $\delta V_{\bf i}$ equal to the volume of the cell it occupies. Hence the marginalised likelihood is approximated by
\begin{equation}
   P({\bf D}|{\bf M_k})\approx\frac{1}{V_k}\sum\limits^{{\bf i}=N}_{{\bf i}=1}P({\bf D}|{\bf M_k},{\bf x_i})\,\delta V_{\bf i}\,.
\end{equation}
Here $N$ is the number of points on the Markov Chain and $V_k$ is the total volume of parameter space associated with model $\bf{M_k}$. As MCMC sampling is most noisy around the edges of the distribution, we omit from the summation any cells that extend to the boundaries of the parameter space. These regions are under sampled and have disproportionately large cells; including them generally overestimates $P(\bf{D}|\bf{M_k})$.

Note that $1/V_k$ is the probability density of the prior PDFs given in Equations (\ref{priorpdf1}) to (\ref{priorpdf4}. This term decreases exponentially with the number of free parameters in each model. So, for example, model {\bf M4} would need to produce a much better fit to the data than {\bf M1} to give $P({\bf D}|{\bf M4})>P({\bf D}|{\bf M1})$.

The relative likelihood of one model, ${\bf k}$, with respect to another, ${\bf k'}$, is quantified by the Bayes factor
\begin{equation}
K_{\bf kk'}=\frac{P(\bf{M_k}|\bf{D})}{P(\bf{M_{k'}}|\bf{D})}=\frac{P(\bf{D}|\bf{M_k})P(\bf{M_k})}{P(\bf{D}|\bf{M_{k'}})P(\bf{M_{k'}})}\,.
   \label{bfactor}
\end{equation}
Given that we have no \emph{a priori} preference for either model, i.e. $P(\mathbf{M_k})=P(\mathbf{M_{k'}})$, Equation (\ref{bfactor}) reduces to the ratio of the marginalised likelihoods. Bayes factors quantifying the relative performance of the different models are presented in Table \ref{bftable}\,.

\section{Results}\label{results}%

\subsection{Parameter estimation for {\bf M1}}\label{paramestm1}%

Figure \ref{m1params} shows the posterior PDFs for $\sigma_{_{\rm O}}$ in {\bf M1}, based on the different data sets. Since the PDFs are all unimodal and not overly skewed, we can calculate means and standard deviations, viz. $\sigma_{_{\rm O}}\!=\!0.63\pm0.08\,({\rm MAN98}),\;0.56\pm0.03\,({\rm SSGK06}),\;0.51\pm0.06\,({\rm SNW08})$.

Also shown in this figure is the product of all three PDFs, i.e. our overall inference on the parameter $\sigma_{_{\rm O}}$ given all three data sets. As the result is reasonably similar across all three data sets, we infer a value of $\sigma_{_{\rm O}}\!=\!0.57\pm0.06$, i.e. the principal axes of a core typically differ by a factor of order $\exp(\sigma_{_{\rm O}})\!\approx\!1.8\pm0.1$.

Figure \ref{aspect_data1} compares the observed distributions of aspect ratio from the different data sets with the best fits from {\bf M1}. {\bf M1} fits the SSGK06 data (which, with 111 core shapes, has the least noisy statistics) well. The fits to the SNW08 data (52 cores) and the MAN98 data (36 cores) are less good. For example, the MAN98 data hints at a sharp peak between $q\!=\!0.5$ and $q\!=\!0.6$ which {\bf M1} is unable to reproduce; however, this may just be a consequence of small-number statistics.

\subsection{Parameter estimation for {\bf M2a}}\label{paramestm2}%

Figure \ref{m2params} shows the posterior PDFs of $\mu_{_{\rm O}}$ and $\sigma_{_{\rm O}}$ in {\bf M2a}, based on the different data sets. We recall that $\mu_{_{\rm O}}$ determines whether cores have a tendency to be prolate ($\mu_{_{\rm O}}\!<\!0$) or oblate ($\mu_{_{\rm O}}\!>\!0$). For all three data sets there is a degeneracy, because the intrinsic asymmetry of the cores is promoted both by increasing $\sigma_{_{\rm O}}$, and by increasing $|\mu_{_{\rm O}}|$. Consequently solutions with reduced $\sigma_{_{\rm O}}$ and increased $|\mu_{_{\rm O}}|$ have high probability. Indeed, for the MAN98 and SSGK06 data-sets these are actually the preferred solutions. However, in neither case is there a clear preference for prolate over oblate cores, or {\it vice versa}. Figure \ref{m2params} also shows the product of the three individual posterior PDFs. This PDF is double peaked, like those of SSGK06 and MAN98, and there is still no preference for cores with prolate or oblate shapes.

Figure \ref{aspect_data2} compares the observed distributions of aspect ratio with the best fits from {\bf M2a}. {\bf M2a} delivers a markedly better fit -- than {\bf M1} -- to the MAN98 and SSGK06 data-sets, irrespective of whether we use the prolate or oblate parameters. However, the best fit to the SNW08 data set is not much better than with {\bf M1}.

\subsection{Parameter estimation for {\bf M2b}}\label{paramestm3}%

Figure \ref{m3params} shows the posterior PDFs of $\sigma_{_{\rm B}}$ and $\sigma_{_{\rm C}}$ in {\bf M2b}, based on the different data sets. For the SSGK06 data we find a peak at $\sigma_{_{\rm B}}\!\approx\!\sigma_{_{\rm C}}\!\approx\!0.55$, and for the MAN98 data at $\sigma_{_{\rm B}}\!\approx\!\sigma_{_{\rm C}}\!\approx\!0.60$. For the SNW08 data, the distribution of $\sigma_{_{\rm B}}$ and $\sigma_{_{\rm C}}$ is somewhat broader, but nowhere does it exceed that for $\sigma_{_{\rm B}}\!\approx\!\sigma_{_{\rm C}}\!\approx\!0.50$. Thus, for all three data sets, the two parameters of {\bf M2b} do not provide a better fit than the single parameter of {\bf M1}.

\subsection{Parameter estimation for {\bf M4}}\label{paramestm4}%

Figure \ref{m4params} shows the posterior PDFs of $\mu_{_{\rm B}}$ and $\mu_{_{\rm C}}$ in {\bf M4}, based on the different data sets. For simplicity, we have marginalized $\sigma_{_{\rm B}}$ and $\sigma_{_{\rm C}}$ out of the PDFs. With the SNW08 data, we see that predominantly oblate cores (i.e. $\mu_{_{\rm B}}\approx0$ and $\mu_{_{\rm C}}<0$, or $\mu_{_{\rm B}}<0$ and $\mu_{_{\rm C}}\approx0$) and triaxial cores with $\mu_{_{\rm B}}\approx\mu_{_{\rm C}}\approx0$) are preferred over prolate cores (i.e. $\mu_{_{\rm B}}\approx\mu_{_{\rm C}}<0$). Conversely, with the SSGK06 and MAN98 data, we see that predominantly prolate cores with $\mu_{_{\rm B}}\approx\mu_{_{\rm C}}\approx-0.4$ and triaxial cores with $\mu{_{\rm B}}\not\approx\mu_{_{\rm C}}\not\approx0$ are preferred. Combining the three PDFs as a product, we see that there is comparable probability for prolate cores with $\mu_{_{\rm B}}\approx\mu_{_{\rm C}}\approx-0.4$, oblate cores with $\mu_{_{\rm B}}\approx0,\mu_{_{\rm C}}\approx-0.8$, and triaxial cores with parameters in the region between these two extremes. We note that these three families of shapes can also be delivered by ${\bf M2a}$ and ${\bf M1}$ with fewer than four parameters.

\citet{GW-TW02} and \citet{JBD01} perform a similar analysis using the core observations catalogued by \citet*{JMA99}. This covers 264 cores from multiple star forming regions and includes cores with embedded protostars (which we do not). Their analysis also differs in that they draw intrinsic aspect ratios from a Gaussian distribution (we use a lognormal distribution in order to avoid negative aspect ratios). \citet{GW-TW02} infer that $\mu{_{\rm B}}=-0.2$ and $\mu{_{\rm C}}=-0.9$ and \citet{JBD01} infer that $\mu{_{\rm B}}=-0.1$ and $\mu{_{\rm C}}=-0.7$\,. We see from the combined posterior PDFs in Figure \ref{m4params} that these values are within the most likely 95\% of the distribution and therefore we are not in direct disagreement. However, the posterior PDF here is only inferred from starless cores in Ophiuchus and permits a large variety of triaxial shapes. 

\subsection{Model selection}%

We quantify the quality of the different models, for the different data sets, by calculating Bayes factors, $K_{\bf kk'}$, as described in Section \ref{selection}. The results are presented in Table \ref{bftable}, where $K>1$ indicates a preference for the model denoted in the column header, and $K<1$ indicates a preference for the model in the row label. \citet*{Jeffreys1961} suggests the following qualitative interpretation for different values of $K_{\bf kk'}$:
\begin{align*}
   &K_{\bf kk'}\le1/10&  \text{Strongly supports }\mathbf{M_{\bf k'}}\,,\\
   1/10<&K_{\bf kk'}\le1/3 & \text{Moderately supports }\mathbf{M_{\bf k'}}\,,\\
   1/3<&K_{\bf kk'}<1 & \text{Weakly supports }\mathbf{M_{\bf k'}}\,,\\
   &K_{\bf kk'}=1 & \text{No preference}\,,\\
   1<&K_{\bf kk'}<3 & \text{Weakly supports }\mathbf{M_{\bf k}}\,,\\
   3\le&K_{\bf kk'}<10 & \text{Moderately supports }\mathbf{M_{\bf k}}\,,\\
   &K_{\bf kk'}\ge10 & \text{Strongly supports }\mathbf{M_{\bf k}}\,;
\end{align*}
we stress that these categories are only intended to be indicative.

The SNW08 data are much better fitted by {\bf M1} or {\bf M2b}, than by {\bf M2a} or {\bf M4}; there is little to chose between {\bf M1} and {\bf M2b}. Conversely, the SSGK06 and MAN98 data sets are both fitted best by {\bf M2a}, with {\bf M1} also giving a good fit, and {\bf M2b} and {\bf M4} giving relatively poor fits. To combine the data sets, we have simply taken the products of their individual Bayes factors, and these are given in the last panel of Table \ref{bftable}. These values suggest that {\bf M1} is the best model. {\bf M2a} is almost as good, and should remain in the reckoning against the day when sufficient data is available to distinguish between prolate and oblate cores. 

\begin{table}
   \begin{center}
      \begin{tabular}{|l|cccc|}
      \hline
      \multicolumn{5}{|c|}{MAN98}\\
      \hline
       & {\bf M1} & {\bf M2a} & {\bf M2b} & {\bf M4} \\
      \hline
      {\bf M1} & 1 & 1.76 & 0.75 & 0.60 \\
      {\bf M2a} & 0.57 & 1 & 0.43 & 0.34 \\
      {\bf M2b} & 1.33 & 2.34 & 1 & 0.80 \\
      {\bf M4} & 1.67 & 2.93 & 1.25 & 1 \\
      \hline
      \multicolumn{5}{|c|}{SSGK06}\\
      \hline
       & {\bf M1} & {\bf M2a} & {\bf M2b} & {\bf M4} \\
      \hline
      {\bf M1} & 1 & 1.80 & 0.31 & 0.48 \\
      {\bf M2a} & 0.56 & 1 & 0.17 & 0.27 \\
      {\bf M2b} & 3.23 & 5.80 & 1 & 1.54 \\
      {\bf M4} & 2.10 & 3.76 & 0.65 & 1 \\
      \hline
      \multicolumn{5}{|c|}{SNW08}\\
      \hline
       & {\bf M1} & {\bf M2a} & {\bf M2b} & {\bf M4} \\
      \hline
      {\bf M1} & 1 & 0.31 & 1.02 & 0.08\\
      {\bf M2a} & 3.20 & 1 & 3.27 & 0.26 \\
      {\bf M2b} & 0.98 & 0.31 & 1 & 0.08\\
      {\bf M4} & 12.11 & 3.78 & 12.36 & 1 \\
      \hline
      \multicolumn{5}{|c|}{\sc Combined}\\
      \hline
       & {\bf M1} & {\bf M2a} & {\bf M2b} & {\bf M4} \\
      \hline
      {\bf M1} & 1 & 0.98 & 0.24 & 0.023 \\
      {\bf M2a} & 1.02 & 1 & 0.24 & 0.024 \\
      {\bf M2b} & 4.21 & 4.21 & 1 & 0.10 \\
      {\bf M4} & 42.5 & 41.6 & 10.0 & 1 \\
      \hline
      \end{tabular}
   \end{center}
   \caption[Bayes factors of shape fitting models.]{Bayes factors, $K\!=\!P({\bf M}_{_{\rm COLUMN}}|{\bf D})/P({\bf M}_{_{\rm ROW}}|{\bf D})$, calculated using Eqn. \ref{bfactor}. The first three panels give values for the individual data sets, and the fourth panel gives their product.}
   \label{bftable}
\end{table}

\subsection{Errors}%

Since our analysis has not included the errors on individual data points (they are not available), the errors in Figures \ref{aspect_data1} and \ref{aspect_data2}, and in Eqn. (\ref{chi2}), should be larger. This would broaden the posterior PDFs for all models, but the effect would tend to be larger for models with more free parameters, in the sense that the probability would be smeared over more dimensions, and therefore their marginal likelihoods would be reduced more. Since we have already concluded that {\bf M1} performs best, we infer that this conclusion would be reinforced if observational errors were included.

\section{Conclusions}\label{conclusions}%

We have used Bayesian analysis to infer the intrinsic shapes of starless cores in Ophiuchus. We find that the observational data are well fitted with a one-parameter model, {\bf M1}, in which cores are triaxial ellipsoids with axes chosen from a log-normal distribution having zero mean and standard deviation $\sigma_{_{\rm O}}\!\approx\!0.57\pm 0.06$. This suggests that the intrinsic axes of cores typically vary by a factor of $F\approx1.8\pm0.1$\,. The two-parameter model {\bf M2b} does not sufficiently improve the fit to justify its adoption, and the four-parameter model, {\bf M4} is completely unjustified.

{There is some evidence to suggest that model {\bf M2a} performs as well as {\bf M1}. However, the strong degeneracy between mostly oblate cores and mostly prolate cores makes it impossible to establish whether either shape is dominant, and this situation may not improve until we have greatly improved observational data, since the projected distributions of $q$ for prolate and oblate cores are very similar. In the meantime, the additional free parameter in model {\bf M2a} does not produce a sufficiently improved fit to justify its use.

Given that {\bf M1} is a simple model with a single well constrained parameter, we will use it in future to define the initial shapes of starless cores. We can randomly draw intrinsic aspect ratios from the model and use sizes taken from observations to fully define the ellipsoidal shapes of cores.}

\section*{Acknowledgements}

We thank the anonymous referee for their helpful feedback. We gratefully acknowledge the support of the STFC, via a doctoral training account (OL) and a rolling grant (OL \& APW; ST/K00926/1). AC gratefully acknowledges the support of a Royal Society Dorothy Hodgkin Fellowship.

\bibliographystyle{mn2e}
\bibliography{refs}
\label{lastpage}
\end{document}